# Capsule and convolutional neural network-based SAR ship classification in Sentinel-1 data

*Leonardo De Laurentiis, Andrea Pomente, Fabio Del Frate, and Giovanni Schiavon*

*University of Rome Tor Vergata, Via del Politecnico, 1, Rome, Italy*

## Abstract

Synthetic Aperture Radar (SAR) constitutes a fundamental asset for wide-areas monitoring with high-resolution requirements. The first SAR sensors have given rise to coarse coastal and maritime monitoring applications, including oil spill, ship and ice floes detection. With the upgrade to very high-resolution sensors in the recent years, with relatively new SAR missions such as Sentinel-1, a great deal of data providing a stronger information content has been released, enabling more refined studies on general targets features and thus permitting complex classifications, as for ship classification, which has become increasingly relevant given the growing need for coastal surveillance in commercial and military segments. In the last decade, several works focused on this topic have been presented, generally based on radiometric features processing; furthermore, in the very recent years a significant amount of research works have focused on emerging deep learning techniques, in particular on Convolutional Neural Networks (CNN). Recently Capsule Neural Networks (CapsNets) have been presented, demonstrating a notable improvement in capturing the properties of given entities, improving the use of spatial informations, in particular of spatial dependence between features, a severely lacking feature in CNNs. In fact, CNNs pooling operations have been criticized for losing spatial relations, thus special capsules, along with a new iterative routing-by-agreement mechanism, have been proposed. In this work a comparison between Capsule and CNNs potential in the ship classification application domain is shown, by leveraging the OpenSARShip, a SAR Sentinel-1 ship chips dataset; in particular, a performance comparison between capsule and various convolutional architectures is built, demonstrating better performances of CapsNet in classifying ships within a small dataset.

## 2. Method

In the Earth Observation field of study, and even more in conducting SAR studies, researchers have to deal very frequently with small datasets. This is mainly due to the fact that very high-resolution SAR products are often commercial and only released under particolar conditions. Thanks to the European Copernicus programme, many Sentinel data have been released under an open-access freely available license. This allows an increased data availability, which represents a fundamental milestone for all the studies that aim to apply deep learning in the Earth Observation (EO) field, particularly for SAR, given the fact that deep learning is well-known to require thousands, sometimes millions of training data. As compared with common datasets generally exploited in deep learning studies, the dataset employed in the present work can be regarded as a small dataset, considering that it is in the order of thousands training samples; as mentioned, this is an extremely common situation in the EO field, therefore it turns out to be very interesting the study of possible architectures and settings capable of high accuracy performances in similar situations. From the results obtained in the present paper, in a small- dataset framework, Capsule Neural Networks are shown to have finest performances in terms of ship classification accuracy as compared to popular deep architectures. A relevant factor in the field of ship detection and classification is represented by the polarization acquisition mode. As detailed in previous works, **[9,19]** vessels and their metallic structures have different and important backscattering behaviours within the different polarization channels. In particular, both for dual-pol and quad- pol acquisition modes, cross-polarization channels are found to perform well for ship detection; moreover, in Sentinel-1 OpenSARShip data, VH polarization is shown to be more sensitive to the changes of velocities and to target structure. The obtained results are then expected to be in accordance with the referred remarks and, as evidenced in the Results section, Sentinel-1 VH polarized data generally lead to an improved ship classification accuracy when compared to Sentinel-1 VV polarized data, as regards CapsNet. At a glance, a benchmark comparison between a CapsNet and various CNNs in terms of ship classification accuracy is shown in this work, demonstrating the CapsNet superiority when dealing with a small dataset, a very common setting in EO field of study. Finally, CNNs performances in a standard data augmentation framework are shown, in order to find out the augmented training data dimensionality required to achieve performances similar to CapsNet; at this stage standard data augmentation operations, such as UpDown-LeftRight flipping and random rotations, are applied. Moreover, by using several GANs, individually trained to learn a specific ship class data distribution, the training dataset undergoes a different data augmentation process that consists in generating data of the different ship classes for the purposes of balancing and uniforming the ship classes size. This process is shown to bring benefits in the classification accuracy, in the same way as the standard data augmentation techniques

## 2.1 Dataset

In this work, the employed dataset has been produced starting from the OpenSARShip, **[9]** a dataset designed to study and improve maritime applications. This latter has been built by leveraging 41 Sentinel-1 products with different environmental conditions, delivering 11346 SAR Sentinel-1 ship tiles, integrated with their respective automatic identification system (AIS) messages. The OpenSARShip dataset delivers two products, in IW mode: Single Look Complex (SLC) and Ground Range Detected (GRD); the dataset built for the present work has been based on GRD products. Sentinel-1 IW default polarization mode, which is the dual-pol VH-VV, is the polarization mode of all ship tiles in the dataset. As stated in the original paper, data are saved using a single 32-bit format, in compliance with the Sentinel-1 original data format. In particular, for GRD products, each ship tile is saved into a matrix which stores amplitude values of pixels for both VH and VV polarizations; furthermore, GRD products delivered by OpenSARShip have been radiometrically calibrated, by leveraging SNAP 3.0; the obtained data, stored in terms of backscatter coefficient σ 0 , constitutes the final data considered to build the dataset employed in this paper. Among all the different ship classes present in OpenSARShip, which covers 17 AIS types, three classes have been selected: tanker, container ship and bulk carrier, since these three ship types represent roughly the 80% of the international shipping market. **[11]**

In particular, in order to build the final dataset, the ElaboratedType and AISShipInformation values in the xml Metadata file (attached to each of the OpenSARShip tiles) have been considered. Specifically, a Container ship is considered to belong to that class if the ElaboratedType corresponds to Container Ship and if the AISShipInformation corresponds to an integer value in a range of 70-79; the same considerations applies to Tanker (AISShipInformation in a range of 80-89) and Bulk Carrier (AISShipInformation in a range of 70-79) classes. The final dataset consists of 2738 ship tiles, and for each of them both VH and VV channels are present; eventually, the entire dataset has been allocated on a 64-16-20 proportion basis, a typical training-validation-test sets division in machine learning; general specifications of the employed dataset are summarized in **Table 1**. Given that there is not a uniform tile size in OpenSARShip, the final dataset has been resized to a common size format of 128x128x1, which constitutes the input format of employed Neural Networks; in particular, a linear interpolation has been applied in case of tiles with dimensions larger than 128x128, whereas a padding, with values equal to the sea background backscattering magnitude, in the other cases. It is noteworthy that the final dataset turns out to be unbalanced, as shown in **Table 2**, therefore GANs have been used as a data-augmentation tool in order to investigate experimental results of CNNs in a rebalanced framework. Specifically, each of the employed GANs specialized in learning a given ship class data distribution, going through a 2000 epochs training stage; once the adversarial training phase ended, the generator model has been used to generate new ship samples, in such a way that the original dataset has been rebalanced to 2000 samples per ship class. By way of illustration, the Generator Loss of the Tanker-specialized GAN is reported in **Figure 1**. A decreasing trend is evident, notably from epoch ∼ 250 onwards, meaning that the Generator network starts to model well the original data distribution, fooling the Discriminator; similar trends take place in training the other specialized GANs, suggesting that Generators are able to produce very-realistic ship tiles. A few generated samples are shown in **Figure 2**, in where it is possibile to observe a general visual comparison with some ship samples from the original dataset.

## 2.3 Neural Networks hyperparameters setting

As said in the previous sections, different neural network architectures have been used in this work. With regard to the employed CNNs, namely VGG-16, VGG-19, Xception, InceptionResNet and NASNet, they have been implemented through Keras, **[26]** a Python-based high-level NNs API, able to run seamlessly on CPUs and GPUs and above various backends such as TensorFlow, CNTK, Theano; in particular, the TensorFlow backend has been selected at this stage. By abstracting out the first layers, that jointly compose the fundamental logic of each network, Keras provides the opportunity to define the fully connected layers at the top of every net. Specifically, all of the CNNs fully connected layers have been aligned to two different patterns: a small 32-16-3 3-layers architecture and a larger 1024-512-3 3-layers architecture, with rectified linear unit activation functions. All networks exploited for ship classification purposes (including the CapsNet) have been trained for 50 epochs, in order to provide uniformity; for CNNs the Adam optimizer has been chosen (learning rate=0.001), with batches size set to 32 and the cross-entropy (also known as log loss) as loss function. The only exception is represented by the NASNet architecture, wherein the batches size has been reduced to 16, due to memory allocation issues. On the other hand, settings and hyperparameters of the CapsNet architecture generally follows the indications given in the original paper. 17 Therefore, a first Conv-layer with 256 9x9 kernels is followed by 32 channels of convolutional 8-dimensional capsules and a final layer with a 16-dimensional capsule per ship class. Adam optimizer with learning rate set to 0.001 and a decay-factor set to 0.9 has been used to minimize the margin losses ($L_k$ in the original paper) sum. Finally, a decoder network has been added for reconstruction purposes, by means of which a sum-of-squared-differences reconstruction loss, scaled by 0.0005, is added to the margin losses. The entire net has been trained for 50 epochs, in order to have same conditions as the CNNs, with batches size set to 100. The GANs specialized in learning a specific ship class data distribution all have the same architecture, and have been trained using Adam optimizer with learning rate=0.002 and $\beta_1$ =0.5, by way of cross entropy loss functions both for discriminator and for the whole generator-discriminator network; furthermore, GANs have been trained for 2000 epochs, with batches size set to 32, before the generation takes place. The generator architecture, shown in **Figure 3**, is composed of a dense fully-connected layer, taking a normally distributed noise vector and whose output is moved on to 4 sequentially deconvolutional layers, all trained with the objective of generating ship tiles resembling the original ones. On the other hand, the discriminator architecture is built in a specular way, with a unique otput value, recalling that its ultimate goal is the prediction of fake-generated or true-original data.

## 3. Results

In this section experimental results are shown, in order to give a comprehensive overview of the different networks performances. In particular, the Capsule Neural Network has been trained on the original dataset built for the present work (described in section 2.1), showing the accuracy scores detailed in **Table 3**. This latter evidently exhibits better performances of the CapsNet when trained and tested on VH-polarized data, as compared to VV-polarized data; furthermore training (and testing) on the complete dataset, including both VH and VV polarized data, is shown to improve the overall accuracy performances. This is in accordance with the results recalled in section 2, claiming cross-polarized channels lead to better ship detection performances and that, particularly in Sentinel-1 OpenSARShip data, VH-polarized data are more sensitive to the changes of velocities and to target structure thus better performances are expected, and found, in the ship classification domain as well. Evidently, training on VH-polarized data leads to refined feature extraction and network training such that it even benefits in the VH-VV case, resulting in higher scores likely related with the bigger dataset dimensionality as well. Moreover, training-validation loss and accuracy trends are shown in **Figure 4**, exhibiting a bigger learning potential in the first epochs, that slowly increases in the subsequent stages.

*Table 3: CapsNet test-set accuracy performances*

| Architecture | VH | VV | VH-VV |
| --- | --- | --- | --- |
| CapsNet | 0.65091 | 0.24363 | 0.66727 |

**Table 4** indicates the accuracy performances of selected CNNs architectures, trained on the original dataset in two different configurations, S and L, that correspond to the above-mentioned fully-connected smaller 32-16-3 3-layers architecture and larger 1024-512-3 3-layers architecture, placed at the end of each CNN for the final classification purposes. Above all, it is very clear that there is no CNN which turns out to be superior to CapsNet in dealing with ship classification through VH and VH-VV data. A different result is evident moving the focus on VV-polarized data; in this case, CNNs prove to be more efficient in extracting features and informations helpful to better classify ships. As a general rule, VH-polarized data don't show a clear advantage over VV-polarized data in the CNN-architectures scenario, except for a few cases, demonstrating that CNNs feature detectors work well both with VH and VV data; it is likely that CapsNets are more sensitive to the lack of unique hierarchical and accurate structures of ships, more confused in VV-polarized data, while CNNs only need the presence of sparse features throughout the tiles in order to properly classify ships. However it turns out that the use of a full VH-VV dataset generally leads to finest accuracy scores even in the CNNs scenario. By all means, the best accuracy score over the original dataset is kept by the CapsNet, trained on the full VH-VV tiles set, which demonstrates its superiority over CNNs in classifying ships on a small dual-pol SAR Sentinel-1 dataset.

*Table 4: CNNs test-set accuracy performances.*

| Architecture | VH | VV | VH-VV |
| --- | --- | --- | --- |
| VGG-16 (S) | 0.35937 | 0.35937 | 0.39705 |
| VGG-16 (L) | 0.37890 | 0.35937 | 0.39705 |
| VGG-19 (S) | 0.35937 | 0.37890 | 0.39705 |
| VGG-19 (L) | 0.35937 | 0.35937 | 0.39705 |
| NASNet (S) | 0.41176 | 0.40808 | 0.40625 |
| NASNet (L) | 0.44117 | 0.45588 | 0.522058 |
| Xception (S) | 0.625 | 0.64843 | 0.64705 |
| Xception (L) | 0.60156 | 0.58984 | 0.62132 |
| InceptionResNet (S) | 0.61718 | 0.62109 | 0.64889 |
| InceptionResNet (L) | 0.64843 | 0.57812 | 0.63786 |

CNNs are well-known to work great with a large amount of training data and, in principle, to improve their performances with the addition of new training data. As a complement to the present study, the accuracy performances of the Xception architecture (selected as one of the two CNNs with top scores in the present framework) are investigated in a data-augmentation context with different augmentation policies, A and B, and reported in **Table 5**. In particular, under the A policy all the training samples undergo flipping operations about horizontal and vertical axes (thus tripling the training set), while under the B policy all the training samples undergo flipping operations about horizontal and vertical axes, as well as a random rotation (hence quadrupling the training set). The results show that by increasing the training dataset by a factor 3 or a factor 4, accuracy performances of the Xception CNN become comparable, even slightly higher, to those of CapsNet. Besides, it is confirmed again that the full VH-VV dataset generally lead to better accuracy performances with respect to the individual VH and VV datasets.

*Table 5: Xception test-set accuracy performances with data-augmentation.*

| Architecture | VH | VV | VH-VV |
|---|---|---|---|
| Xception (S) (A) | 0.69531 | 0.65625 | 0.67647 |
| Xception (L) (A) | 0.63281 | 0.63281 | 0.70772 |
| Xception (S) (B) | 0.64062 | 0.61328 | 0.68198 |
| Xception (L) (B) | 0.625 | 0.67578 | 0.68566 |

CNNs performances are ultimately evaluated into a GANs data-augmentation context, in order to analyse potential benefits given by the use of generated ship samples. Within this framework, trained GANs are then exploited with the aim of balancing and increasing training data dimensionality; specifically, in this study GANs have been used to balance the ship samples to the value of 2000 tiles per ship class (see **Table 2**). Results reported in **Table 6** demonstrate that the outlined data-augmentation approach has the potential to improve accuracy performances, at least for the top-two networks; in fact, those two are led to scores comparable to the one achieved by CapsNet in the VH-VV scenario.

*Table 6: Xception test-set accuracy performances with data-augmentation.*

| Architecture | VH-VV |
|---|---|
| VGG-16 (L) | 0.39705 |
| VGG-19 (L) | 0.39705 |
| NASNet (L) | 0.42095 |
| Xception (L) | 0.65625 |
| InceptionResNet (L) | 0.66544 |

## 4. Conclusions and further developments

In this work CapsNet and various CNNs performances are investigated in the ship classification application domain, through the use of a dataset based on the OpenSARShip, a SAR Sentinel-1 ship dataset. Within this small dataset framework, CapsNet demonstrate to have finest capabilities in classifying ships, by harnessing VH-polarized data and dual VH-VV-polarized data; on the other hand, CNNs demonstrate higher performances in using VV-polarized data. As reported in other research works, VH polarization results to be more sensitive to changes of velocities and to target structures in Sentinel-1 OpenSARShip data, hence it is possible that CapsNet is more sensitive to the lack of unique hierarchical and accurate structures of ships, that appear to be more confused in VV-polarized data, while CNNs only need the presence of given sparse features throughout the tiles in order to properly classify ships, thus resulting in better accuracy performances. Nevertheless, the best accuracy score over the dataset built for the present work is obtained by the CapsNet, trained on the full VH-VV tiles set, demonstrating its superiority in classifying ships over a small dual-pol SAR Sentinel-1 dataset. Standard data augmentation techniques are then examined, in order to find out if a larger dataset can lead CNNs to results comparable with the CapsNet, and factors 3 and 4 are found sufficient for the Xception architecture. Finally, CNNs performances are evaluated in a different data-augmentation framework; in this last, each of a set of GANs are specialized in learning a given ship class data distribution, and then their generators are used to produce synthesized ship samples, with a view to rebalance and augment the original dataset. This latter approach demonstrates to have a good potential to improve accuracy performances; in fact, the top-two networks are led to scores comparable to the one achieved by CapsNet in the VH-VV scenario, hence the GAN-approach could be further investigated in order to examine if the addition of new generated samples corresponds to ulterior accuracy improvements.